\input psfig
\documentstyle[12pt,ws-art-j]{article}
\renewcommand{\today}{\number\day\space \ifcase\month\or January\or February\or
March\or April\or May\or June\or July\or August\or September\or October\or
November\or December\fi \space\number\year .}
\begin{document}
\title{ STUDY OF CHIRAL INVARIANCE USING PIONS\thanks {Work supported
by the United States National Science Foundation.}
	}

\author{DINKO PO\v{C}ANI\'C \\
{\em Department of Physics, University of Virginia,} \\
{\em Charlottesville, VA, 22901-2458, U.S.A.} \\
{\em e-mail: pocanic@virginia.edu }
        }
\maketitle
\vskip -5mm
\begin{abstract}
We review two manifestations of chiral symmetry breaking in the pion sector.
First, we examine the $\pi\pi$ scattering amplitude at threshold, which is
completely determined by the chiral symmetry breaking part of the strong
interaction, and, thus, directly constrains the form of the low energy
effective lagrangians.  We review the current status of the $\pi\pi$
scattering lengths in view of the recent results on reactions $\pi N
\rightarrow\pi\pi N$ near threshold, and discuss the needed improvements.
Second, we focus on the rate of the pion beta decay, determined at the tree
level by the CVC hypothesis, a remnant of chiral symmetry breaking.  Precise
knowledge of the $\pi\beta$ decay rate constrains
Cabibbo--Kobayashi--Masakawa unitarity, and, thus, can be used to distinguish
among certain extensions of the minimal Standard Model. Present status of
CKM unitarity, and a new precise measurement of the $\pi\beta$ decay rate at
PSI are discussed.
\end{abstract}

\section {Introduction: Chiral Symmetry and Pions}

Chiral transformations are the most general form of rotations in fermion
flavor space; for free massless fermions they leave the lagrangian
invariant.  Thus, the simplest realization of chiral symmetry is in a
massless two-fermion system. Pure ``isospin'' rotations in the
two-dimensional flavor space of fermions ``$u$'' and ``$d$'',
$$ \psi' = \left(\matrix{u'\cr d'\cr}\right)=
\exp(i\vec{\alpha}\cdot\vec{\tau})\left(\matrix{u\cr d\cr}\right) ,
\eqno{(1)}$$
leave the free massless lagrangian unchanged, forming an SU(2)$\times$U(1)
symmetry group.  The generators, $\vec{\tau}$, are the Pauli spin--1/2
matrices (the U(1) generator is the 2$\times$2 unit matrix).  This, however,
is not the maximal symmetry of the massless two-fermion system.  Independent
rotations in flavor space of projected left-- and righthanded states,
$$ \matrix{\psi_{_L}={1\over 2}(1-\gamma_5)\psi = \left(\matrix{u_{_L}\cr
             d_{_L}\cr }\right) & \quad \psi_L'=
             \exp(i\vec{\alpha}_L\cdot\vec{\tau})\psi_L , \cr
   \psi_{_R}={1\over 2}(1+\gamma_5)\psi = \left(\matrix{u_{_R}\cr
             d_{_R}\cr }\right) & \quad \psi_R'=
             \exp(i\vec{\alpha}_R\cdot\vec{\tau})\psi_R , \cr} \eqno{(2)} $$
form the full chiral symmetry group of the lagrangian, U(1)${\rm
_L}\times$SU(2)${\rm _L}\times$SU(2)${\rm _R}\times$U(1)${\rm _R}$;
$\gamma_5$ is the familiar Dirac matrix.  Usually, attention is focused on
the nonabelian SU(N)${\rm _L}\times$SU(N)${\rm _R}$ subgroup.  Strong
interactions leave the chiral SU(2)${\rm _L}\times$SU(2)${\rm _R}$ symmetry
for the $u$ and $d$ quarks nearly intact, whereas the SU(3)${\rm
_L}\times$SU(3)${\rm _R}$ group, obtained by including the $s$ quark, is
violated considerably more.  This is not surprising given that fermion mass
terms, $-m\bar{\psi}\psi$, break chiral invariance of the lagrangian by
mixing the left-- and righthanded states, and that $m_s\gg m_{u,d}$.

For hadrons, chiral symmetry is broken in a very particular way: (i)
explicitly, through small non-zero quark masses, and (ii) ``spontaneously'',
through the non-invariance of vacuum to chiral transformations, but only to
their axial-vector part, leading to the well known ``partially conserved
axial-vector current'' (PCAC) relation.  The effect of spontaneous chiral
symmetry breaking (ChSB) dominates and hides the full symmetry from the
casual observer; at the same time, it is the source of a rich phenomenology,
requiring the existence of an octet of pseudoscalar Goldstone bosons,
$\pi$'s, $K$'s, and $\eta$. Clearly, chiral symmetry plays a major role in
the strong dynamics at low energies, one that has been much studied
throughout the past four decades.\cite{quig83}

Chiral symmetry has recently attained a new significance due to the failure
of the full QCD to describe nonperturbative strong interaction phenomena
below about 1 GeV.  At these energies chiral symmetry provides the basic
framework or the essential constraints for all low energy effective
lagrangian models, e.g., the chiral perturbation theory
(ChPT),\cite{gasser84} and recent QCD formulations of the Nambu--Jona-Lasinio
(NJL) model in SU(2) and SU(3) flavor space.\cite{klev92}

We examine two aspects of chiral invariance in the pion sector: (i) the
mechanism of explicit ChSB, as evidenced in threshold $\pi\pi$ scattering,
and (ii) repercussions of the spontaneous ChSB on weak-interaction physics,
in particular the pion beta decay and its implications regarding the Standard
Model.

\section {Pion-Pion Scattering and the Explicit Breaking of Chiral Symmetry}

It has been long established\cite{weinb66} that low energy $\pi\pi$
scattering provides a particularly sensitive tool in the study of the explicit
breaking of chiral symmetry, since $a(\pi\pi)$, the $\pi\pi$ scattering
lengths, vanish exactly in the chiral limit.  To the extent that they differ
from zero, $a(\pi\pi)$ provide a direct measure of the ChSB term in the pion
sector, i.e., detailed knowledge of the $\pi\pi$ scattering lengths and phase
shifts provides much needed input in fixing the parameters of ChPT and other
effective models.

In the 1960's, before the advent of QCD, $\pi\pi$ scattering lengths were
studied in order to distinguish between the various proposed dynamical
mechanisms of ChSB in strong interactions.  Weinberg's,\cite{weinb66}
Schwinger's\cite{schwin67} and Chang and G\"ursey's\cite{chang67} effective
lagrangians differ from each other in the form of the response of the pion
field to chiral transformations, thus predicting different values for the
s-wave $\pi\pi$ scattering lengths. (It was subsequently established that
Weinberg's lagrangian has the leading-order ChSB term consistent with QCD.)

Weinberg's soft-pion prediction was improved in 1982 by Jacob and
Scadron who introduced a correction due to the non-soft
$S^*\rightarrow\pi\pi$ isobar background.\cite{jacob82}  Soon thereafter,
Gasser and Leutwyler applied chiral perturbation theory to the problem, and
predicted the behavior of the $\pi\pi$ amplitude to order
$p^4$.\cite{gasser83} Also inspired by QCD, Ivanov and Troitskaya used the
model of dominance by quark loop anomalies (QLAD) to obtain $\pi\pi$
scattering lengths.\cite{ivanov86} The J\"ulich group, however, constructed a
meson exchange model of pseudo\-scalar-pseudo\-sca\-lar meson scattering,
and applied it to calculate a number of $\pi\pi$ and $K\pi$ scattering
observables.\cite{lohse90} Calculations of $\pi\pi$ scattering lengths within
the SU(2)$\times$SU(2) and SU(3)$\times$SU(3) realizations of the NJL model
are found in Refs.~\cite{ruivo94} and \cite{berna91}, respectively. Kuramashi
and coworkers successfully applied quenched lattice QCD on a 12$^3\times$20
lattice, and obtained $\pi\pi$ scattering length values in the
neighborhood of the older current-algebra calculations.\cite{kuram93}
Roberts {\it et al.} recently developed a model field theory,
referred to as the global color-symmetry model (GCM), in which the
interaction between quarks is mediated by dressed vector boson
exchange.\cite{rober94}  Predictions  of all these models are compared to
experimental results below.

\subsection {Experimental Determination of $\pi\pi$ Scattering Lengths}

Experimental evaluation of $\pi\pi$ scattering observables is difficult,
primarily because free pion targets are not available.  Scattering lengths
further require measurements close to the $\pi\pi$ threshold, where phase
space strongly limits measurement rates.  Several reactions have been studied
or proposed as a means to obtain near-threshold $\pi\pi$ phase shifts: $\pi
N\rightarrow \pi\pi N$, $K_{e4}$ decays, $\pi^+\pi^-$ atoms,
$e^+e^-\rightarrow \pi\pi$, etc.  Only the first two reactions have so far
proven useful in studying threshold $\pi\pi$ scattering.  The three main
methods are discussed below in the order of decreasing reliability.

\subsubsection {$K_{e4}$ Decays}

The $K^+\rightarrow\pi^+\pi^- e^+\nu$ decay (called $K_{e4}$) is in several
respects the most suitable process for the study of near-threshold $\pi\pi$
interactions.  The interaction takes place between two real pions on the mass
shell, the only hadrons in the final state.  The dipion invariant mass
distribution in $K_{e4}$ decay peaks close to the $\pi\pi$ threshold, and
$l=I=0$ and $l=I=1$ are the only dipion quantum states contributing to the
process.  These factors, and the well understood $V-A$ nature of the weak
decay, favor the $K_{e4}$ process among all others in terms of theoretical
uncertainties.  On the other hand, measurements are impeded by the low
branching ratio of the decay, $\sim 3.9\times 10^{-5}$.

The most recent $K_{e4}$ measurement was made by a Geneva--Saclay
collaboration in the mid-1970's who detected some 30,000 $K_{e4}$
decays.\cite{rosse77}   Analysis of this low-background, high-statistics data
illustrates well the difficulties encountered in extracting experimental
$\pi\pi$ scattering lengths. By itself, the Geneva--Saclay experiment
determines $a_{\ell=0}^{I=0}$ with a $\sim$35\% uncertainty, and constrains
the slope parameter $b$:\cite{rosse77}
$$ a_0^0 = 0.31 \pm 0.11\ \mu^{-1} ,\hskip 5mm b = b_0^0 - a_1^1 = 0.11
\pm 0.16 \ \mu^{-1} , \eqno (3)$$
reflecting the relative insensitivity of $K_{e4}$ data
to $a(\pi\pi)$.  Pion-pion phase shifts are, however,
constrained by unitarity, analyticity, crossing and Bose symmetry, which were
extensively studied by Roy,\cite{roy71} and Basdevant {\it et
al.},\cite{basde72,basde74} resulting in a set of relations known as the
``Roy equations''.  By using the Roy equation constraints of Basdevant, {\it
et al.},\cite{basde74} more accurate results for $a_0^0$ and $b_0^0$ were
obtained,\cite{rosse77} as well as scattering lengths and slope parameters
for ($l=0$, $I=2$) and ($l=1$, $I=1$).\cite{nagels79}

\subsubsection{Peripheral $\pi N \rightarrow \pi \pi N$ Reactions: the
Chew--Low Method}

Particle production in peripheral collisions can be used to extract
information on low-energy scattering of two of the particles in
the final state, as shown by Chew and Low in 1959.\cite{chew59}  For
the $\pi N \rightarrow \pi \pi N$ reaction, the Chew--Low formula,
$$
\sigma_{\pi\pi}(m_{\pi\pi}) = \lim_{t\rightarrow\mu^2}\ \left[
{\partial^2\sigma_{\pi\pi N} \over \partial t\partial m_{\pi\pi} } \cdot {\pi
\over \alpha f_\pi^2} \cdot {p^2(t-\mu^2)^2 \over tm_{\pi\pi}k} \right],
\eqno (4) $$
relates $\sigma_{\pi\pi}(m_{\pi\pi})$, the cross section for pion-pion
scattering, to double differential $\pi N \rightarrow \pi\pi N$ cross section
and kinematical factors: $p$, momentum of the incident pion, $m_{\pi\pi}$,
the dipion invariant mass, $t$, the Mandelstam square of the 4-momentum
transfer to the nucleon, $k=(m_{\pi\pi}^2/4-\mu^2)^{1/2}$, momentum of the
secondary pion in the rest frame of the dipion, $f_\pi \approx 93$ MeV, the
pion decay constant, and $\alpha=1$ or 2, a statistical factor involving the
pion and nucleon charge states.  The method relies on an accurate
extrapolation of the double differential cross section to the pion pole in
order to isolate the one pion exchange (OPE) pole term contribution.  Since
the exchanged pion is off-shell in the physical region ($t<0$), this method
requires measurements under conditions which maximize the OPE contribution
and minimize all background contributions, i.e., peripheral pion production
with $t$ as close to zero as possible, and incident momenta typically above
$\sim$ 3 GeV/c.

The Chew--Low method has been refined considerably over time, particularly by
Baton and coworkers,\cite{baton70} whose approach enables extraction of
$\pi\pi$ phase shifts through appropriate treatment of the angular dependence
of the $\pi N\rightarrow \pi\pi N$ exclusive cross sections.  The ``Roy
equation''\cite{roy71,basde72,basde74} dispersion constraints enable the use
of accurate high energy peripheral pion production data, whereas such data
are not available below $m_{\pi\pi} \approx 500$ MeV.  The peripheral pion
production data base has not essentially changed since the early 1970's, and
is dominated by two experiments, performed by the Berkeley\cite{proto73} and
CERN-Munich\cite{grayer74} groups.  These data were ultimately combined with
the Geneva--Saclay $K_{e4}$ data in a comprehensive dispersion-relation
analysis,\cite{nagels79} as discussed in the preceding section.

There have been independent Chew--Low type analyses since the 1970's; the
last published one, performed by the Kurchatov Institute group in 1982,
was based on a set of some 35,000 events recorded in bubble
chambers.\cite{aleks82}  The same group has recently updated their
analysis.\cite{patar93}

\subsubsection{$\pi N \rightarrow \pi \pi N$ Reactions near Threshold}

Weinberg pointed out early on that the OPE graph plays an important role in
the $\pi N \rightarrow \pi \pi N$ reactions near threshold.\cite{weinb66}
Olsson and Turner's soft-pion model incorporates only the OPE and contact
terms in the lagrangian at threshold,\cite{olss68} leading to a
straightforward relation between the $\pi\pi$ and $\pi N \rightarrow \pi \pi
N$ threshold amplitudes with only one parameter, $\xi$. Thus, soft-pion
theory has provided the main inspiration for the $\pi N \rightarrow \pi \pi
N$ near-threshold measurements.  In Olsson--Turner's model, it is sufficient
to measure total $\pi N \rightarrow \pi \pi N$ cross sections, from which
quasi-amplitudes can be calculated and extrapolated to threshold.
Unfortunately, the Olsson--Turner parameter $\xi$ is incompatible with QCD.

Unlike the two preceding methods, there has recently been a great deal of
experimental activity on near-threshold $\pi N \rightarrow \pi\pi N$
measurements.  There are 5 charge channels accessible to measurement,
$\pi^-p\rightarrow\pi^-\pi^+n$, $\pi^-p\rightarrow\pi^0\pi^0n$,
$\pi^-p\rightarrow\pi^-\pi^0p$, $\pi^+p\rightarrow\pi^+\pi^0p$, and
$\pi^+p\rightarrow\pi^+\pi^+n$.  We label them for brevity with their final
state charges as ($-$+n), ..., (++n), respectively.  Recent experiments
reporting total $\pi N\rightarrow\pi\pi N$ cross sections are summarized
below, while data available before 1984 is reviewed in Ref.~\cite{manley84}

The OMICRON group at CERN has measured cross sections in the (+$-$n),
($-$0p), (++n) and (+0p) channels by detecting coincident charged particle
pairs in a two-sided in-plane magnetic spectrometer.\cite{kernel89} Sevior
and coworkers at TRIUMF measured inclusive cross sections for the reaction
(++n) down to 5 MeV above threshold, using a novel technique involving an
active plastic scintillator target and neutron detection.\cite{sevior91}
J. Lowe and coworkers at Brookhaven measured the channel (00n)down to about 5
MeV above the threshold, using the Crystal Box detector, in a kinematically
complete measurement.\cite{lowe91}  Finally, a Virginia-Stanford-LAMPF team
studied the (+0p) channel in an exclusive measurement using the LAMPF $\pi^0$
spectrometer and an array of plastic scintillation telescopes for $\pi^+$ and
$p$ detection.\cite{pocan94,frlez93}  Several other experiments are
currently under way at TRIUMF and LAMPF, aiming to bring the existing cross
section data base even closer to the $\pi\pi$ threshold.

Apart from a discrepancy in the (++n) channel between the
TRIUMF\cite{sevior91} and OMICRON\cite{kernel89} data, the entire body of
data appears globally consistent within the quoted
uncertainties.\cite{burk91,pocan94,frlez93} We also note the high-statistics
angular correlation measurements in the (+$-$n) channel performed by the
Erlangen group at somewhat higher energies than the bulk of the other work in
this Section.\cite{ortner90} Closer analysis of the exclusive cross sections
is only now beginning, as high-statistics data were not available until
recently.  Therefore, the methods of analysis are still being refined. Recent
work of interest has been reported by the Erlangen\cite{jaek90} and St.
Petersburg\cite{boloh91} groups, who also discuss earlier work of other
groups not included in this brief review.

\subsection{Comparison of Experimental Results and Predictions for
$a(\pi\pi)$}

\begin{figure}[hbt]
\vbox{\centerline{\psfig{figure=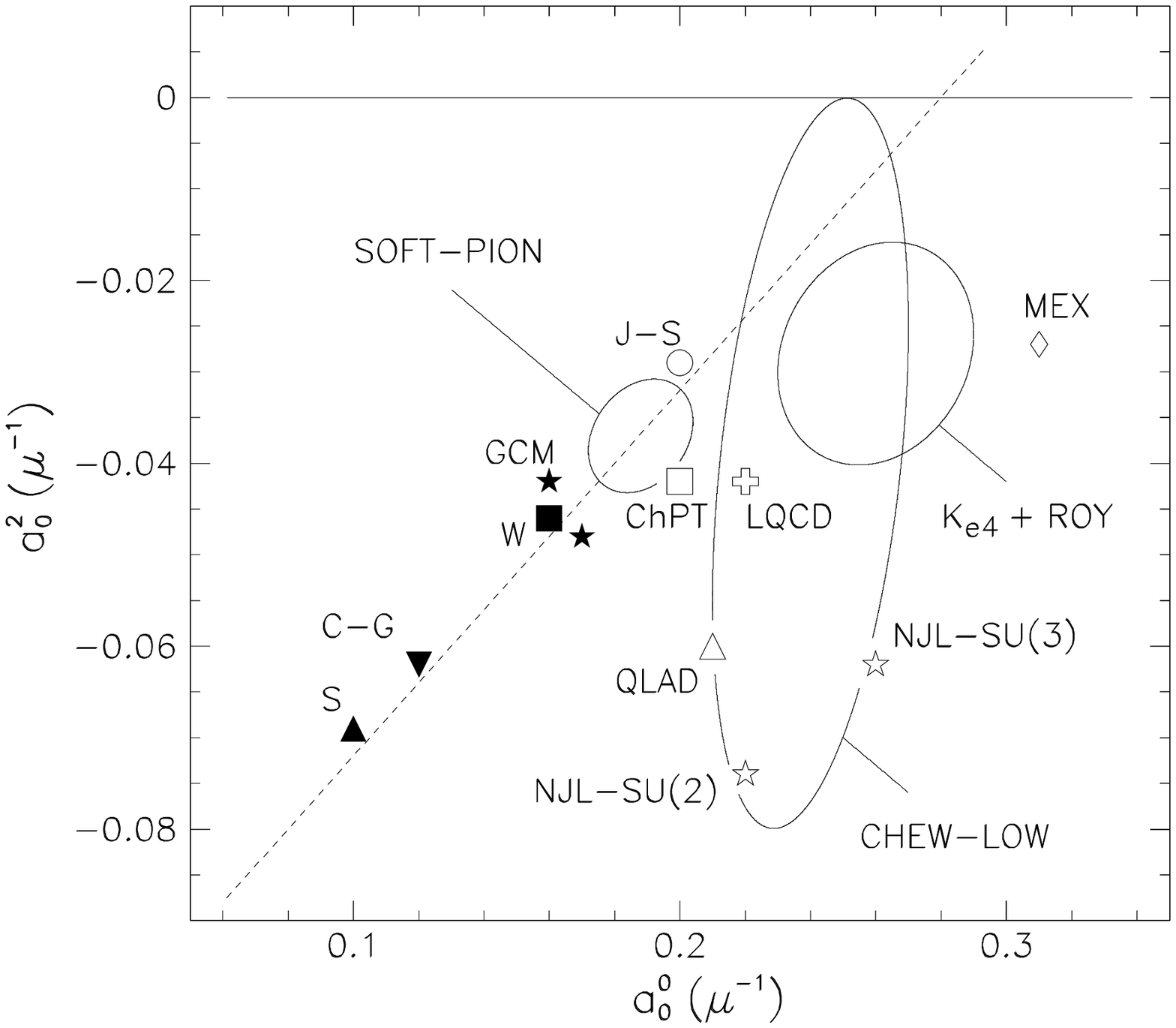,height=100mm}} }
\caption{ \small  \baselineskip=12pt
Summary of the $\pi\pi$ scattering length predictions (symbols)
and experimental results (contour limits).
Model calculations: Weinberg (full square), Schwinger (filled
triangle), Chang and G\"ursey (filled inverted triangle), Jacob and Scadron
(open circle), Gasser and Leutwyler -- ChPT (open square), Ivanov and
Troitskaya -- QLAD (open triangle), Lohse {\it et al.} -- Meson Exchange
(open rhomb), Ruivo {\it et al.} and Bernard {\it et al.} -- NJL (open
stars), Kuramashi {\it et al.} -- quenched lattice gauge QCD (open cross),
Roberts {\it et al.} -- GCM (filled stars).
\label{f1} }
\end{figure}

Figure~\ref{f1}. summarizes quantitatively the theoretical model
predictions and the experimental determinations of the s-wave $\pi\pi$
scattering lengths.

Considerable scatter of predicted values of $a_0^0$ and $a_0^2$ is evident in
Fig.~\ref{f1}.  Even disregarding the 1960's calculations of
Schwinger\cite{schwin67} and Chang and G\"ursey\cite{chang67} later
superseded by QCD, a significant range of predicted values remains.  However,
in view of the present experimental uncertainties (see Fig.~\ref{f1}.), most
authors claim that their results are supported by the data.  This is not a
satisfactory state of affairs. Progress is required on two fronts in order to
improve the present uncertainties in the experimentally derived $\pi\pi$
scattering lengths.

First, more accurate data on $K_{e4}$ decays are needed, in order to reduce
the current error limits in the analysis.  However, $K_{e4}$ data alone will
not suffice because of their insensitivity to the $I=2$ $\pi\pi$ channel.
Measurements of the $\pi^+\pi^-$ atom proposed at several laboratories, if
feasible with reasonable statistics, could provide the needed additional
information.  The quantity to be measured is the decay rate of the
$\pi^+\pi^-$ ground state into the $\pi^0\pi^0$ channel, which is
proportional to $|a_0^0-a_0^2|^2$.\cite{urets61}

Second, the mounting volume and accuracy of the exclusive and inclusive
near-threshold $\pi N\rightarrow \pi\pi N$ data contain, in principle,
valuable information regarding threshold $\pi\pi$ scattering.  Theoretical
interpretation of this data requires much improvement in order to make full
use of this information.  The work of Bernard, Kaiser and
Mei{\ss}ner\cite{berna94} appears to be a promising step in that direction.

\section {CVC, the Pion Beta Decay, and Extensions to the Standard Model}

Numerous hadronic processes, besides $\pi\pi$ scattering, are dominated by
chiral symmetry at low energies; however, to appreciate fully the
implications of chiral symmetry, we turn our attention to the weak
interactions of hadrons.  Since weak interactions do not conserve hadronic
flavor, ChSB effects are pronounced in weak decays and transitions.

A linear transformation changes the ($L$,$R$) fermion states to the ($V$,$A$)
basis,
$$ \matrix{\psi_V= {1\over 2}(\psi_R+\psi_L) \quad , \quad &
           \psi_V'= \exp(i\vec{\alpha}_V\cdot\vec{\tau})\psi_V \quad ,\quad &
              \alpha_V^a={1\over 2}(\alpha_R^a+\alpha_L^a) \cr
           \psi_A= {1\over 2}(\psi_R-\psi_L) \quad , \quad &
           \psi_A'= \exp(i\vec{\alpha}_A\cdot\vec{\tau})\psi_A \quad, \quad &
              \alpha_A^a={1\over 2}(\alpha_R^a-\alpha_L^a).\cr}
  \eqno (5) $$
As mentioned in Sect.~1, unlike axial-vector transformations, vector
rotations in hadronic flavor space leave the vacuum invariant, preserving a
chiral symmetry subgroup SU(N)$_V\times$U(1)$_V$ which is associated, through
Noether's theorem, with a conserved vector current (CVC).  Early recognition
of this symmetry, along with the observation that the coupling constants in
leptonic and hadronic beta decays are nearly the same, led to the concept of
quark-lepton (Cabibbo) universality.  Study of pure vector (Fermi) beta decays
to date has confirmed the CVC hypothesis to a high degree.  An interesting
aspect of pure vector semileptonic transitions
is their sensitivity to the physics beyond the Standard Model through the
Cabibbo--Kobayashi--Masakawa (CKM) unitarity.

\subsection {Motivation for a Precise Determination of the $\pi\beta$ Decay
Rate}

Pion beta decay, $\pi^\pm\rightarrow\pi^0e^\pm\nu$, is one of the fundamental
semileptonic weak interaction processes.  It is a transition between two
spin-zero members of an isospin triplet, and is therefore analogous to
superallowed pure Fermi transitions in nuclear beta decay.  Due to its
accuracy of theoretical description, Fermi beta decay stands today among the
prominent successes of the Standard Model (SM) of electroweak interactions.

The rate of pure Fermi nuclear beta decays is directly related to the vector
coupling constant $G_V^2$ through the CVC hypothesis. Two types of
nucleus-dependent corrections are needed in order to extract $G_V$ from
measured decay rates: (a) phase space dependent radiative corrections
$\delta_R$ due to electroweak and QCD loop diagrams, and (b) nuclear
structure dependent corrections $\delta_C$.

Radiative corrections of $O(\alpha)$ and the leading terms of $O(Z\alpha^2)$
have been calculated, while contributions of $O(Z^2\alpha^3)$ have been
estimated.\cite{Sir-87a}  The resulting $\delta_R$'s amount to 3--4\%,
typically,  with a theoretical uncertainty of 0.2\% arising mainly from
short-distance axial-vector-induced contributions.\cite{Mar-86}.

On the other hand, the nuclear corrections $\delta_C$\ are smaller, ranging
from 0.2 to 1\%, but their evaluation appears to be accompanied by
significant ambiguities.\cite{Orm-85} Past evaluations of $\delta_C$ have
yielded inconsistent results\cite{Tow-77,Wil-78,Orm-85} The discrepancy was
recently removed by Hardy {\it et al.}\cite{Har-90} Using their result, one
obtains $|V_{ud}|^2 + |V_{us}|^2 + |V_{ub}|^2$ = 0.9962 $\pm$ 0.0016, i.e., a
value 2.4$\sigma$ lower than the SM prediction.

$V_{ud}$ can also be determined from neutron beta decay using $\tau_n$ and
$G_A/G_V$.  The current world average of the two quantities yields $V_{ud}$ =
0.9804 $\pm$ 0.0022, bringing $|V_{ud}|^2 + |V_{us}|^2 + |V_{ub}|^2$ up to
1.0096 $\pm$ 0.0044, which also differs from the SM prediction by
2.2$\sigma$, but in the {\it opposite} direction.

Thus, the determination of $V_{ud}$ and the test of universality is presently
not in a satisfactory state at the level of a few tenths of a percent.
Determination of the $\pi^\pm\rightarrow\pi^0e^\pm\nu$ rate at the same level
is, therefore, important.  Already at about 0.5\%, such a measurement would
provide an independent test of universality in a meson.  It is of particular
interest to find out where the pion beta decay result will fall, whether it
will be consistent with the nuclear Fermi decays, or with the neutron decay
data.

Pion beta decay is, furthermore, free of nuclear screening and overlap
corrections $\delta_C$, and involves a simpler
axial-vector contribution to the radiative correction $\delta_R$.\cite{Mar-86}
The overall theoretical uncertainty involved in evaluating $G_V^2$ or
$|V_{ud}|^2$ from an experimental $\pi\beta$ decay rate is thus $\leq 0.1\%$
\cite{Mar-86,Sir-89}  The only other significant source of uncertainty in
such an evaluation involves the pion mass difference $\Delta = m_{\pi^+} -
m_{\pi^0}$.  $\Delta$ has been recently determined\cite{Cra-88} to an
accuracy of 10$^{-4}$, and thus contributes just 0.05\% towards a total
uncertainty of $\leq 0.11\%$.  The theoretical advantage is, however, offset
by experimental difficulties due to the small $\pi\beta$ branching ratio of
$\sim 10^{-8}$; hence the present experimental accuracy of 4\%.\cite{McF-85}
The following physics becomes accessible through a precise constraint on the
CKM unitarity:

     (a) A possible fourth fermion generation could manifest itself through
$u$--$b'$ mixing inferred from an experimental violation of three-generation
CKM unitarity.\cite{Mar-86}  Present data give only an upper bound of $V_{ub'}
\le 0.07$ at the 90\% confidence limit.

      (b) Another possible source of quark-lepton universality violation
involves supersymmetric particles, as postulated in SUSY extensions to the
minimal SM.  Barbieri {\it et al.} have pointed out that superparticle
exchanges in $\mu$ and $\beta$ decay give rise to a violation of quark-lepton
universality, unless their masses are suitably constrained.\cite{Bar-85}
Significant regions in the ($m_{\widetilde q}$, $m_{\widetilde \ell}$,
$m_{\widetilde W}$) space can be excluded on the basis of presently available
superallowed Fermi nuclear $\beta$ decay data.  For example, given reasonable
assumptions on the higgsino mass parameter range, i.e., $m_{\widetilde h} \ge
O$(100~GeV), the combination $m_{\widetilde q}$ = 80 GeV, $m_{\widetilde
\ell}$ = 25 GeV, $m_{\widetilde W}$ = 30 GeV is excluded.

      (c) Possible additional neutral gauge bosons $Z'$, besides the known
$\gamma$ and $Z$ of the standard electroweak theory, would also affect CKM
unitarity.  Marciano and Sirlin have computed quantum loop corrections to the
quark mixing angles resulting from several types of $Z'$.\cite{Mar-87}
Present data infer lower bounds on $m_{Z'}$ ranging from 254 to 413 GeV,
depending on the choice of the GUT model for the $Z'.$\cite{Sir-89,Mar-87}

      (d) Compositeness of elementary fermions and vector gauge bosons
presents a possible way to extend the SM.  Sirlin has examined the
consequences of several models of compositeness on electroweak radiative loop
corrections.\cite{Sir-87b} In this framework, present experimental limits on
the validity of quark-lepton universality can be used to constrain the mass
scale of compositeness to the range above $\Lambda \ge 7$ TeV.\cite{Sir-89}

Above constraints are comparably restrictive to those extracted so far from
direct high-energy measurements.  Thus, the present level of theoretical
accuracy in the interpretation of vector $\beta$ decay processes enables a
selective exploration of certain phenomena on a mass scale largely
inaccessible to existing accelerators.\cite{Mar-87,Sir-87b,Sir-89} A new,
more precise measurement of the pion beta decays is particularly well suited
to extending further the above limits on new physics.

\subsection{The PIBETA experiment at PSI}

A new project is under way at the Paul Scherrer Institute (PSI) with the aim
of improving the present accuracy of the pion beta decay rate by an order of
magnitude.  This objective is being pursued in stages; the goal of the first
phase is a 0.5\% measurement.  The experiment is a collaboration of seven
institutions: the University of Virginia, PSI, Arizona State University, INS
Swierk, JINR Dubna, Tbilisi State University and the Rudjer Bo\v{s}kovi\'c
Institute.\cite{Pocan92}

The PIBETA experiment relies on detection of pion decays at rest.  An
intense low energy $\pi^+$ beam ($1-2\times 10^6 \pi$/s) is stopped in a
fragmented active target made up of thin scintillating fibers.  The signature
of $\pi\beta$ decays is given by the pair of back-to-back photons following
$\pi^0$ decay.  The photons are detected in a segmented 240-module pure CsI
electromagnetic shower calorimeter subtending 3/4 of 4$\pi$.  The calorimeter
is spherical in shape, made up of truncated irregular hexagonal and regular
pentagonal pyramids.  The thickness of the pure CsI modules is 22 cm, or 12
radiation lengths.  The inner radius of the CsI sphere is 26 cm, and the
outer 48 cm.  Light produced by showers is read out by photomultiplier tubes
with quartz windows. The associated positron in $\pi\beta$ decay is also
detected inside the active target.

\begin{figure}[htb]
\vbox{\centerline{\psfig{figure=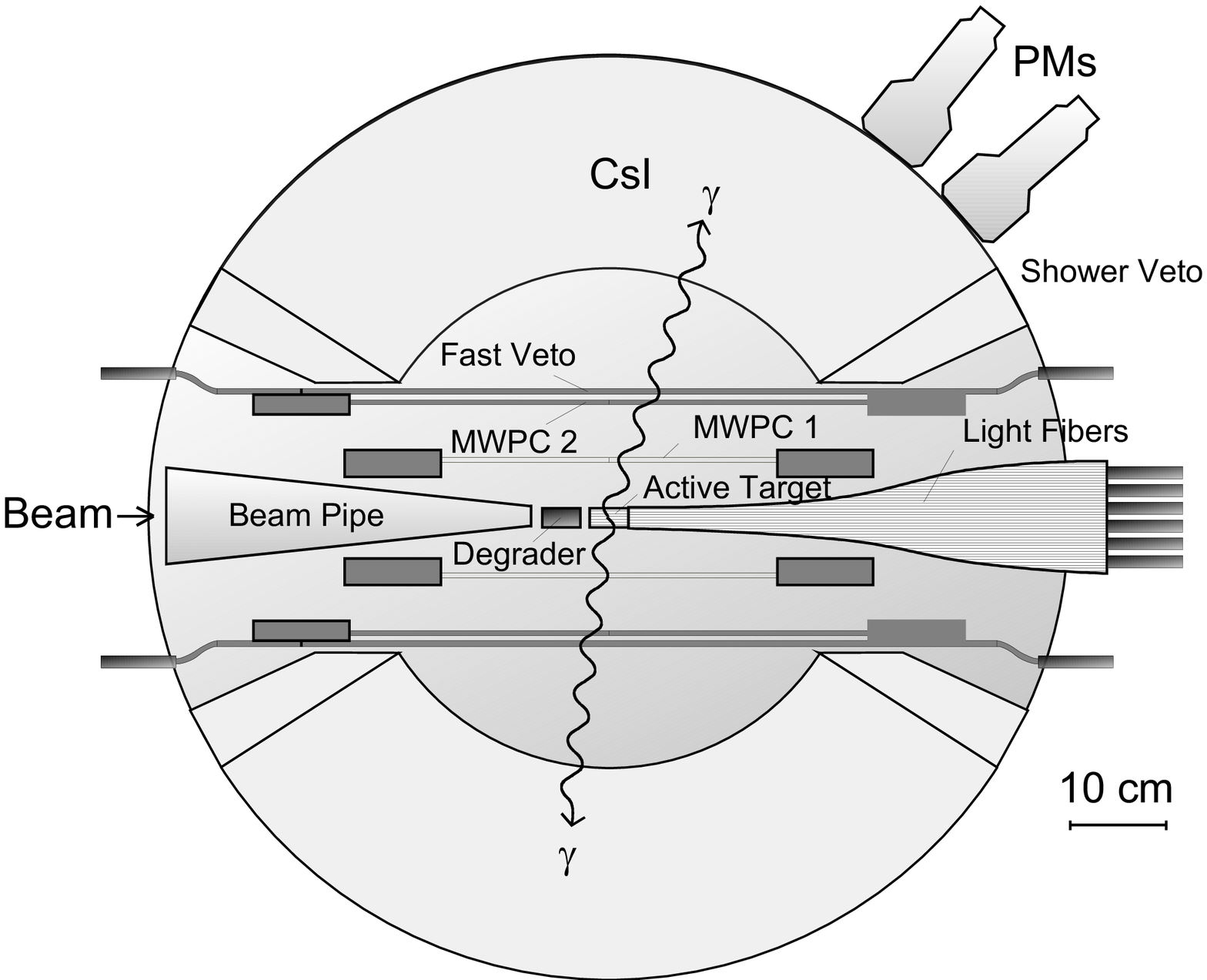,height=120mm}} }
\caption{ \small  \baselineskip=12pt
Schematic cross section of the PIBETA apparatus at PSI.}
\label{f2}
\end{figure}

Measurement of pion beta decays at rest allows the normalization to
$\pi\rightarrow e\nu$ decays with nearly the same systematics (acceptance,
efficiency, energy resolution).  Present accuracy of the $\pi\rightarrow
e\nu$ decay branching ratio, $\simeq$0.3\%,\cite{bern93} suffices for the
first phase of the PIBETA project.  The $e\nu$ signal involves only one
monochromatic positron-induced shower and, therefore, requires a more
demanding trigger.  By far the most important backgroud in both $\pi\beta$
and $e\nu$ detection is presented by the numerous Michel positron events
coming from the decay of muons in the active target, $\mu\rightarrow
e\nu\bar{\nu}$.  In order to separate the various classes of events cleanly,
there are two concentric cylindrical MWPCs for tracking in the central
detector region.  Also, an array of thin plastic scintillator hodoscope
detectors, placed concentrically around the MWPCs serve to identify charged
particles and provide accurate timing.  Figure~\ref{f2} shows the schematic
cross section of the PIBETA detector with all of its main components.  The
entire detector system will be enclosed inside a temperature and humidity
stabilized container, surrounded by and array of cosmic muon veto counters.

In a series of development runs since 1992, the PIBETA collaboration has
addressed and solved the outstanding problems: pion stopping density, muon
and positron contaminations in the beam (held to below 15\% each), pure CsI
surface treatment and energy resolution (currently about 4.2 MeV FWHM for 70
MeV showers), the fast analog shower clustering required for the main trigger
modes, the $\pi\rightarrow e\nu$ trigger, etc.  Quality control for the pure
CsI detector modules (uniformity and percentage of fast component in the
light output) is performed in a dedicated 3-dimensional cosmic muon
tomography setup.

Plans call for the detector system to be assembled and completed by the end
of 1995, while data taking runs are scheduled to begin in 1996.  Due to the
high required measurement precision, accumulating the necessary number of
events will take about 10$^7$ s of running, or most of a calendar year in
real time.  With all the necessary systematic checks and calibration
measurements, it will take several years after detector commissioning to
complete the first, 0.5\% phase.  The second phase, which is to increase the
accuracy by another factor of 2--3, will require instrumental improvements,
and will follow the completion of the first phase.

\section {Conclusions}

The brief discussion of just two aspects of chiral symmetry in the pion sector
serves to illustrate several points pertinent to this subject of study.
First, chiral invariance is intimately related to the pion dynamics at low
energies; indeed, the pion's very existence is tied to chiral symmetry trough
the Goldstone mechanism.  Given the central role chiral dynamics plays at low
energies, the pion sector offers a clean testing ground for effective strong
lagrangians.  Second, details of the spontaneous ChSB also affect
fundamentally the weak interactions of hadrons, making accessible certain
high energy aspects of the Standard Model physics through very precise low
energy measurements.  All of these considerations continue to make the pion
a rich source of information about some of the most basic aspects of
the strong and weak interactions, even after more than three decades of
research.


\begin{thebibliography}{99}

\baselineskip=12.8pt \parsep=0pt \parskip=0pt \itemsep=0pt

\bibitem{quig83} For a comprehensive discussion of ChSB see, e.g., C. Quigg,
{\it Gauge Theories of the Strong, Weak and Electromagnetic Interactions},
(Addison-Wesley, Redwood City, 1983).


\bibitem{gasser84} J. Gasser and H. Leutwyler, {\it Ann. Phys. (N.Y.)}
{\bf 158} (1984) 142; {\it Nucl. Phys.} {\bf B250} (1985) 465.

\bibitem{klev92} A comprehensive recent review of the NJL models can be found
in S. P. Klevansky, {\it Rev. Mod. Phys.} {\bf 64} (1992) 649.

\bibitem{weinb66} S. Weinberg, {\it Phys. Rev. Lett.} {\bf 17} (1966) 616;
{\it ibid.} {\bf 18} (1967) 188.

\bibitem{schwin67} J. Schwinger, {\it Phys. Lett.} {\bf 24B} (1967) 473.

\bibitem{chang67} P. Chang and F. G\"ursey, {\it Phys. Rev.} {\bf 164} (1967)
1752.

\bibitem{jacob82} R. Jacob and M. D. Scadron, {\it Phys. Rev.} {\bf D25}
(1982) 3073.

\bibitem{gasser83} J. Gasser and H. Leutwyler, {\it Phys. Lett.} {\bf 125B}
(1983) 325.

\bibitem{ivanov86} A. N. Ivanov and N. I. Troitskaya, Yad. Fiz. {\bf 43}
(1986) 405 [Sov. J. {\it Nucl. Phys.} {\bf 43} (1986) 260].

\bibitem{lohse90} D. Lohse, J. W. Durso, K. Holinde and J. Speth, {\it Nucl.
Phys.} {\bf A516} (1990) 513.

\bibitem{ruivo94} M. C. Ruivo {\it et al.}, {\it Nucl. Phys.} {\bf A575}
(1994) 460.

\bibitem{berna91} V. Bernard {\it et al.}, {\it Phys. Lett.} {\bf B253}
(1991) 443.

\bibitem{kuram93} Y. Kuramashi 
{\it et al.}, {\it Phys. Rev. Lett.} {\bf 71} (1993) 2387.

\bibitem{rober94} C. D. Roberts 
{\it et al.}, {\it Phys. Rev.} {\bf D49} (1994) 125.

\bibitem{rosse77} L. Rosselet {\it et al.}, {\it Phys. Rev.} {\bf D15}
(1977) 574.

\bibitem{roy71} S. M. Roy, {\it Phys. Lett.} {\bf35B} (1971) 353.

\bibitem{basde72} J. L. Basdevant, J. C. Le Guillou and H. Navelet, {\it
Nuovo Cim.} {\bf 7A} (1972) 363.

\bibitem{basde74} J. L. Basdevant, C. G. Froggatt and J. L. Peterson, {\it
Nucl. Phys.} {\bf B72} (1974) 413.

\bibitem{nagels79} M. M. Nagels {\it et al.}, {\it Nucl. Phys.} {\bf B147}
(1979) 189.

\bibitem{chew59} G. F. Chew and F. E. Low, {\it Phys. Rev.} {\bf 113} (1959)
1640.

\bibitem{baton70} J. B. Baton, G. Laurens and J. Reignier, {\it Phys. Lett.}
{\bf 33B} (1970) 525.

\bibitem{proto73} S. D. Protopopescu {\it et al.}, {\it Phys. Rev.} {\bf D7}
(1973) 1279.

\bibitem{grayer74} G. Grayer {\it et al.}, {\it Nucl. Phys.} {\bf B75} (1974)
189.

\bibitem{aleks82} E. A. Alekseeva {\it et al.}, {\it Zh. Eksp. Teor. Fiz.}
{\bf 82} (1982) 1007. 

\bibitem{patar93} O. O. Patarakin and V. N. Tikhonov, Kurchatov Institute of
Atomic Energy preprint IAE-5629/2 (1993).

\bibitem{olss68} M. G. Olsson and L. Turner, {\it Phys. Rev. Lett.} {\bf 20},
(1968) 1127; {\it Phys. Rev.} {\bf 181} (1969) 2141, L. Turner, {\it Ph.D.
Thesis}, Univ. of Wisconsin, 1969 (unpublished).

\bibitem{manley84} D. M. Manley, {\it Phys. Rev.} {\bf D30} (1984) 536.

\bibitem{kernel89} G. Kernel,  {\it et al.},
{\it Phys. Lett.}  {\bf B216} (1989) 244; {\it ibid.} {\bf B225} (1989) 198;
{\it Z. Phys.} {\bf C48} (1990) 201; {\it ibid.} {\bf 51} (1991) 377; in
{\it Particle Production Near Threshold}, Nashville, 1990, ed. by H. Nann and
E. J. Stephenson, (AIP, New York, 1991).


\bibitem{sevior91} M. E. Sevior {\it et al.}, {\it Phys. Rev. Lett.} {\bf 66}
(1991) 2569.

\bibitem{lowe91} J. Lowe {\it et al.}, {\it Phys. Rev.} {\bf C44} (1991) 956.

\bibitem{pocan94} D. Po\v{c}ani\'c {\it et al.}, {\it Phys. Rev. Lett.}
{\bf 72} 1156 (1994).

\bibitem{frlez93} E. Frle\v{z}, {\it Ph.D. Thesis}, Univ. of Virginia, 1993
(Los Alamos Rep. LA-12663-T).

\bibitem{ortner90} H.-W. Ortner {\it et al.}, {\it Phys. Rev. Lett.} {\bf 64}
(1990) 2759; R. M\"uller {\it et al.}, {\it Phys. Rev.} {\bf C48} (1993) 981.

\bibitem{burk91} H. Burkhardt and J. Lowe, {\it Phys. Rev. Lett.} {\bf 67}
(1991) 2622.

\bibitem{jaek90} O. J\"akel, H.-W. Ortner, M. Dillig and C. A. Z.
Vasconcellos, {\it Nucl. Phys.} {\bf A511} (1990) 733; O. J\"akel, M. Dillig
and C. A. Z. Vasconcellos, {\it ibid.} {\bf A541} (1992) 675; O. J\"akel and
M. Dillig, {\it ibid.} {\bf A561} (1993) 557.

\bibitem{boloh91} A. A. Bolokhov, V. V. Vereshchagin and S. G. Sherman, {\it
Nucl. Phys.} {\bf A530} (1991) 660.

\bibitem{urets61} J. R. Uretsky and T. R. Palfrey, {\it Phys. Rev.} {\bf 121}
(1961) 1798.

\bibitem{berna94} V. Bernard, N. Kaiser and Ulf-G. Mei{\ss}ner, preprint
hep-ph/9404236 (1994).

\bibitem{Sir-87a} A. Sirlin, {\it Phys. Rev. D} {\bf 35} (1987) 3423.

\bibitem{Mar-86} W. J. Marciano and A. Sirlin, {\it Phys. Rev. Lett.} {\bf 56}
(1986) 22.

\bibitem{Orm-85} W. E. Ormand and B. A. Brown, {\it Nucl. Phys.} {\bf A440}
(1985) 274; {\it Phys. Rev. Lett.} {\bf 62} (1989) 866.

\bibitem{Tow-77} I. S. Towner, J. C. Hardy and M. Harvey, {\it Nucl. Phys.}
{\bf A284} (1977) 269.

\bibitem{Wil-78} D. H. Wilkinson, in {\it ``Nuclear Physics with Heavy Ions
and Mesons''}, ed. by R. Balian, M. Rho and G. Ripka (North Holland,
Amsterdam, 1978), p. 877.

\bibitem{Har-90} J. C. Hardy, {\it et al.},
{\it Nucl. Phys.} {\bf A509} (1990) 429.

\bibitem{Sir-89} A. Sirlin, {\sl private communication}, 1989.

\bibitem{Cra-88} J. F. Crawford, {\it et al.},
{\it Phys. Lett.} {\bf 213B} (1988) 391.

\bibitem{McF-85} W. K. McFarlane, {\it et al.},
{\it Phys. Rev. D} {\bf 32} (1985) 547.

\bibitem{Bar-85} R. Barbieri, {\it et al.},
{\it Phys. Lett.} {\bf 156B} (1985) 348; {\it Nucl. Phys.} {\bf B269} (1986)
253.

\bibitem{Mar-87} W.J. Marciano and A. Sirlin, {\it Phys. Rev. D} {\bf 35}
(1987) 1672.

\bibitem{Sir-87b} A. Sirlin, {\it Comments Nucl. Part. Phys.} {\bf 17}
(1987) 279.

\bibitem{Pocan92} D. Po\v{c}ani\'c {\it et al.}, PSI experimental proposal
R-89.01, May 1992.

\bibitem{bern93} G.Czapek, {\it et al.}, {\it Phys. Rev. Lett.} {\bf 70}
(1993) 17.

\end{thebibliography}
\end{document}